\documentclass[vecphys]{svmult}

\usepackage{makeidx}         
\usepackage{graphicx}        
\usepackage{multicol}        
\usepackage[bottom]{footmisc}

\makeindex             

\newcommand{\apj}{\textit{ApJ}}
\newcommand{\aj}{\textit{AJ}}

\newcommand{\apjs}{\textit{ApJ Suppl.}}

\newcommand{\mnras}{\textit{MNRAS}}

\begin{document}

\title*{Properties of voids in the Local Volume}
\author{Anton Tikhonov\inst{1}\and
Anatoly Klypin\inst{2}}
\authorrunning{Local Voids}
\institute{ Department of Mathematics
and Mechanics, St.Petersburg State University
\texttt{avt@gtn.ru}
\and Astronomy Department, NMSU \texttt{aklypin@nmsu.edu}}
%
%
\maketitle

\begin{abstract}

Current explanation of the overabundance of dark matter subhalos in
the Local Group (LG) indicates that there maybe a limit on mass of a
halo, which can host a galaxy. This idea can be tested using voids in
the distribution of galaxies: at some level small voids should not
contain any (even dwarf) galaxies.  We use observational samples
complete to $M_B=-12$ with distances less than 8\,Mpc to construct
the void function (VF): the
distribution of sizes of voids empty of any galaxies. There are $\sim
30$ voids with sizes ranging from 1 to 5\,Mpc.  We also study the
distribution of dark matter halos in very high resolution simulations
of the LCDM model. The theoretical VF matches the observations
remarkably well only if we use halos with circular velocities larger
than $45\pm 10$\,km/s.  This agrees with the Local Group
predictions.   Small voids look quite similar to their
giant cousins: the density has a minimum at the center of a void and
it increases as we get closer to the border.  Thus, both the Local
Group data and the nearby voids indicate that isolated halos below
$45\pm 10$\,km/s must not host galaxies and that small (few Mpc) voids
are truly dark.

\end{abstract}

\section{Introduction}
The observational discovery of giant voids was soon followed by the
theoretical understanding that voids constitute a natural outcome of
structure formation via gravitational instability. Emptiness of voids
-- the number of small galaxies in the voids -- is an interesting
question for both the observations and the theory to tackle
\cite{Peebles2001}. Cosmological simulations predict (e.g.,
\cite{Gottloeber2003}) that many small DM halos should reside in
voids. There seems to be no disagreement between the LCDM theory and
the observations regarding the giant voids defined by $M_*$ galaxies
or by $10^{12}M_{\odot}$ halos \cite{Patiri2006}. The situation is
less clear on smaller scales. In the region of $\sim 10$\,Mpc around
the Milky Way, where observations go to remarkably low luminocities,
small voids look very empty: dwarf galaxies do not show a tendency to
fill the voids and voids are still relatively large.  The theory
predicts that many dwarf dark matter halos should be in the voids,
which puts it in the collision course with observations.  Yet, below
some mass the halos are expected to stop producing galaxies inside
them. There are different arguments for that: stellar feedback
\cite{Dekel1986} or photoionozation may play significant role in
quenching star formation in too small halos. Still, it is difficult to
get a definite answer because the physics of dwarfs at high redshifts
is quite complicated.

Satellites of the Local Group give a more definite answer. Current
explanation of the overabundance of the dark matter subhalos
\cite{Kravtsov2004} assumes that dwarf halos above $V_c \approx
50$\,km/s were forming stars before they fall into the Milky Way or
M31. Once they fall in, they get severely striped and may
substantially reduce their circular velocity producing galaxies such
as Draco or Fornax with the rms line-of-sight velocities only few
km/s.  The largest subhalos retain their gas and continue form stars,
while smaller ones may lose the gas and become dwarf
spheroidals. Halos below the limit never had substantial star
formation. They are truly dark.  This scenario implies that
$V_c\approx 50$\,km/s is the limit for star formation in halos. If this
picture is correct, it can be tested with small-size voids: they must
be empty of any galaxies
 and are filled with gas and dark matter halos.

Tully \cite{Tully1987} noted that the Local
Supercluster contains a number of filaments and that those outline the
so-called Local Void, which begins just outside the Local Group and
extends in the direction of the North Pole of the LSC. The Local Void
looks practically free from galaxies. Over the past few years special
searches for new nearby dwarf galaxies have been undertaken using
numerous observational data. At present, the sample of galaxies with
distances less than 10\,Mpc lists about 500 galaxies. For half of them
the distances have been measured to an accuracy as high as 8-10\%
\cite{Karachentsev2004}. Over the last 5\,years snapshot surveys with
Hubble Space Telescope (HST) have provided us with the TRGB distances
for many nearby galaxies. The absence of the ``finger of God'' effect
in the Local Volume simplifies the analysis of the shape and
orientation of nearby voids. Observations of the Local Volume have
detected dwarf systems down to extremely low luminosity. This gives us
unique possibility to detect voids which may be empty of any
galaxies. Tikhonov \& Karachentsev \cite{Tikhonov2006} analyzed nearby
voids. Here we continue the analyzis using an updated list of galaxies
 (Karachentsev, private communication).  The volume limited sample is
 complete for galaxies with abs. magnitudes $M_B=-12$ within 8\,Mpc
 radius.

We use N-body simulations done with the Adaptive Refinement Tree code
\cite{Kravtsov1997}.  The simulations are for spatially flat
cosmological LCDM model with following parameters: $\Omega_0 = 0.7,
\Omega_{\Lambda} = 0.3; \sigma_8 = 0.9; H_0 = 70$ km/s/Mpc.  As a
measure of how large is a halo we typically use the maxumum circular
velocity $V_c$, which is easier to relate to observatons as compared
with the virial mass.  For reference, halos with $V_c =50$\,km/s have
virial mass about $10^{10}M_{\odot}$ and halos with $V_c =20$\,km/s
have virial mass about $10^{9}M_{\odot}$.  We use two simulations: (1)
Box 80Mpc/h (Box80); mass per particle $3\times 10^8h^{-1}M_{\odot}$;
simulations cover the whole volume and (2) Box 80Mpc/h (Box80S);
 spherical region of 10\,Mpc inside 80Mpc/h box resolved with
 $5\times 10^6h^{-1}M_{\odot}$ particles.

In order to detect voids, we place a 3d mesh on the observational or simulation volume.
We then find initial centers of voids as the mesh centers having the largest distances to
nearest objects. In the next iteration, an initial spherical void may be increased by
adding additional off-center empty spheres with smaller radius. The radius of the spheres
 is limited to be larger then 0.9 of the initial sphere and their centers must stay
inside the volume of the first sphere. The process is repeated few times. It produces
voids which are slightly aspherical, but voids never become more flattened than 1:2 axial ratio.
Artificial objects are palced on the boundaries of the sample to prevent voids getting
 out of the boundaries of the sample.
We define the cumulative void function (CVF) as the fraction of the total volume occupied
 by voids with effective radius larger than
$R_{\rm eff} = (3 V_{\rm void}/4\pi)^{-1/3}$.

\section{Results}\label{sec:cvf}

We use two samples to
construct CVF of the Local Volume: (1) Galaxies
brighter than $M_B=-12$ inside sphere of radius 8\,Mpc and (2) all
galaxies inside 7.5\,Mpc. Results are present in the right bottom panel of
Figure 1. There are about 30 voids in the observational sample. We
limit the radius of voids to be more than 1\,Mpc. The two subsamples
indicate some degree of stability: inclusion of few low-luminocity
galaxies does not change the void function.

We use the Box80 simulation (full volume) to  constract a sample of 40  ``Local Volumes''.
The selecton criteria are:  (1) no
halos with $M > 10^{14}M_{\odot}$ inside a  8\,Mpc sphere (thus, no clusters
in a sample); (2) the sphere must be centered on a halo with $150 < V_c  < 200$
km/s (Milky Way analog); (3)The number of halos found inside 8\~Mpc sphere with $V_c>180$\,km/s
 within 10\% is the average number expected for a sphere of this radius.
 The halo catalogs
 are complete down to halos with circular velocity $V_c=40$\,km/s.
The second simulation (Box80S) provides one sample and it is complete down to 20\,km/s.
The left bottom panel in the Figure~1 shows CVF for different samples of
halos and the observed CVF.  Results indicate that voids in the
distribution of halos with $V_c > 45$\,km/s give the best fit to the observed CVF.
The theoretical  CVF goes above the observational
data if we use circular velocities larger than 60\,km/s. If we use
significantly lower limits, than the theory predicts too few large
voids. The theoretical results match the observations
if we use $V_{circ} = 45 \pm 10$\,km/s. In this case, the match is
 remarkably good: the whole spectrum of voids is reproduced by the theory.

\begin{figure}[htb!]
 \centering
 \includegraphics[scale=0.55]{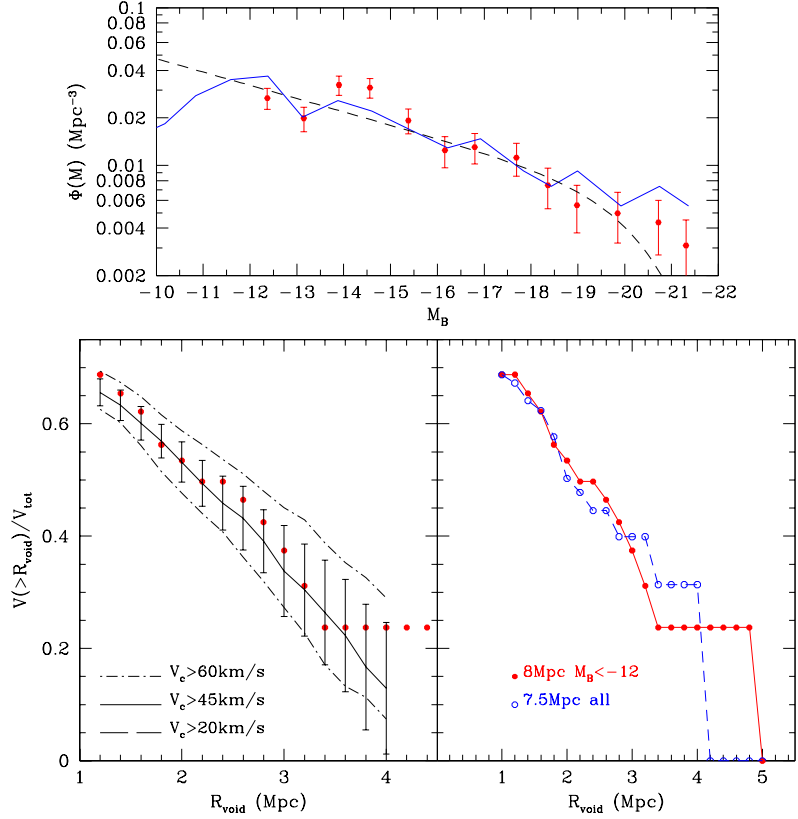}
  \caption{{\bf Bottom Right panel:} the void function for two
  observational samples. The full curve and filled circles are for a
  complete volume limited sample with $M_B<-12$ and $R<8$\,Mpc.  The
  open circles are for all observed galaxies inside
  7.5\,Mpc. Comparison of the samples shows reasonable stability of
  the void function. {\bf Bottom Left panel:} Observational data (the
  complete sample) are compared with the distribution of voids in
  samples of halos with different limits on halo circular
  velocity. CVF for $V_c=45$\,km/s provides a remarkably good fit to
  observations. Note that the LCDM model predicts very large empty
  regions.  {\bf Top panel:} Luminocity function of galaxies. Circles
  with erros show results for 8\,Mpc sample.  The full curve is for
  4\,Mpc sample scaled down by factor 2.7. The dashed curve is for the
  Schechter approximation }
\label{fig:rms} \end{figure}

According to LCDM simulations totally empty front part of the Local
Void is probable. In a sample  of ten 8\,Mpc "Local Volumes" a half of
cases have voids comparable to the largest voids in LV if we consider entire
LV sample.

The top panel in Figure~1 shows the luminocity function of
galaxies. We also show the Schechter approximation with parameters:
$\alpha=-1.21$, $M_*=-19.9+5\log h$, $\Phi_*=1.9\times
10^{-2}h^3$\,Mpc$^{-3}$.  The approximation is the average luminosity
function of galaxies in the B-band in the Universe (not in our
sample).  It provides a very good fit to our data. This means that the
sphere of radius 8\,Mpc containts the average number of galaxies:
$N_{\rm 8Mpc sample}/N_{\rm average} =1$.  We also show the luminosity function for another
complete sample: $R<4$\,Mpc. In this case, the {\it shape} of the
luminosity function is the same, but its normalization is different:
$N_{\rm 4Mpc sample}/N_{\rm average} =2.7$.

\section{Conclusions}\label{sec:concl}

\begin{itemize}
\item The LCDM model is consistent with the cumulative
volume functions of voids in the distribution of galaxies for a large
luminosity range. According to LCDM, large empty voids in Local Volume
such as the Local Void are probable.

\item There are significant (up to few Mpc) holes in the distribution
of halos predicted by LCDM that are free from haloes with $V_c >
20$\,km/s: any haloes of astronomical interest.

\item Voids in the distribution of haloes with $V_c > 45 \pm 10$\,km/s
reproduce the Cumulative Void Function of Local Volume galaxy sample.
We can treat this value as a limit of appearance of a galaxy in a DM
halo.

\item The luminosity function in the Local Volume (8\,Mpc) has the
shape and the normalization of the average LF in the Universe. There
is substantial overdensity of galaxies inside sphere of radius
4\,Mpc. It has 2.7 time more galaxies than the average expected for a
sphere of this size.

\end{itemize}

We thank I.D.\,Karachentsev for providing us an updated list
of his Catalog of Neighboring galaxies. A.\,Klypin acknowledges support of NSF grants to NMSU.
 Computer simulations used in this research
were conducted on the Columbia supercomputer at the NASA Advanced
Supercomputing Division and at the Leibniz-Rechenzentrum (LRZ),
Munchen, Germany. A.\,Tikhonov acknowledges support of grant no.
MK-6899.2006.2 from the President of Russia.


%
%
%
%
%
%
%
%
%
%


\printindex
\end{document}